\documentclass[10pt,aps,prl,twocolumn,showpacs,floatfix]{revtex4-1}

\usepackage{graphicx}
\usepackage{amssymb}
\usepackage{amsmath}
\usepackage{xspace}
\usepackage{color}
\usepackage{braket}

\begin{document}

\title{Spectral and entanglement properties of the bosonic Haldane insulator  
}

\author{Satoshi Ejima}
\author{Florian Lange}
\author{Holger Fehske}
\affiliation{Institut f\"ur Physik, Ernst-Moritz-Arndt-Universit\"at
Greifswald, 17489 Greifswald, Germany}

\date{\today}

\begin{abstract}
We discuss the existence of a nontrivial topological phase in one-dimensional interacting systems described by the 
extended Bose-Hubbard model with a mean filling of one boson per site. Performing large-scale density-matrix renormalization group  calculations we show that the presence of nearest-neighbor repulsion enriches the ground-state phase diagram  of the paradigmatic 
Bose-Hubbard model by stabilizing a novel gapped insulating state, the so-called Haldane insulator, which, embedded into superfluid,  
Mott insulator and density wave phases, is protected by the lattice inversion symmetry. The quantum phase transitions between
the different insulating phases were determined from the central charge via the von Neumann entropy. 
The Haldane phase reveals a characteristic four-fold 
degeneracy of the entanglement spectrum. We finally demonstrate that the intensity maximum of the dynamical charge structure factor, 
accessible by Bragg spectroscopy, features the gapped dispersion known from the spin-1 Heisenberg chain.    
\end{abstract}

\maketitle
A quarter-century after Haldane's conjecture
of an appearance of finite gap in the integer-spin
chain~\cite{Haldane83}, the so-called Haldane phase 
protected by the lattice inversion symmetry attracts renewed 
attention from a topological point of view. Such a topological protected
state, characterized by symmetries and a finite bulk gap, 
is termed now as a symmetry-protected topological (SPT) ordered phase~\cite{GW09,PBTO12}.
In higher dimensions, the so-called Kane-Mele 
topological band insulator of noninteracting fermions~\cite{KM05,KM05b} exhibits a SPT state  
protected by $U(1)$ and time-reversal symmetries. Since particles in real materials
normally interact, it is not sufficient to study SPT order for 
non-interacting systems. To analyze SPT states in interacting systems two main 
approaches have been proposed. The first is based on the definition of 
appropriate topological invariants within a Green function scheme~\cite{Gurarie11}. It has been successfully 
applied to the one-dimensional (1D) Peierls--Hubbard model~\cite{MENG12,YPFK14}.
The second uses the entanglement spectrum as a fingerprint of
topological order~\cite{LH08}. Here the lowest entanglement
level reflects the degree of degeneracy corresponding to
symmetries and the edge states of the system. This has been worked out  
for various spin chains~\cite{PTBO10,PBTO12,LYHYW12}.

Interestingly a hidden SPT phase
was also found in interacting boson systems with long-range
repulsion~\cite{DBA06}. This phase resembles
the Haldane gapped phase of the quantum spin-1 Heisenberg chain.
Indeed, assuming that the site occupation of an 1D extended Bose-Hubbard model
(EBHM) with nearest-neighbor interaction is restricted to $n_j=0, 1$ or~2,
the system can be described by an effective spin-1 model with $S_j^z=n_j-\rho$
for a mean boson filling factor $\rho=1$.
The Haldane insulator (HI) then appears
between the conventional Mott insulator (MI) and
the density wave (DW) phases at intermediate couplings~\cite{DBA06,BDGA08}.
Field theory predicts the MI-HI transition to be in the Luttinger liquid
universality class with central charge $c=1$, whereas the HI-DW transition
belongs to the Ising universality class with $c=1/2$~\cite{BDGA08}. Very recent quantum
Monte Carlo simulations~\cite{BSRG13} reveal in addition a supersolid phase competing
with the HI.

In this work, we focus on the characterization of the EBHM's ground-state and spectral properties from an entanglement point of view. 
Using the (dynamical) density-matrix renormalization group (DMRG) technique~\cite{Wh92,Je02b}, we show that the lowest entanglement level
in the nontrivial topological HI phase is four-fold degenerate.
%, which is reflective of the gapped phase in the spin-1 Heisenberg chain. 
The universality classes of the MI-HI and HI-DW transitions are determined from the central charge in accordance
with what is obtained from field theory. Most notably we demonstrate that the dynamical charge structure factor 
can be used to unambiguously  discriminate the HI from the MI and DW phases.  

The Hamiltonian of the EBHM is defined as
\begin{eqnarray}
\hat{{\cal H}}&=&
 -t\sum_j( \hat{b}_j^\dagger \hat{b}_{j+1}^{\phantom{\dagger}}
          +\hat{b}_{j}^{\phantom{\dagger}}\hat{b}_{j+1}^\dagger )
 +U\sum_j \hat{n}_j(\hat{n}_j-1)/2 \nonumber \\
&& +V\sum_j \hat{n}_j\hat{n}_{j+1},
\label{hamil}
\end{eqnarray}
where $\hat{b}_j^{\dagger}$, $\hat{b}_j^{}$, and  $\hat{n}_j=\hat{b}_j^\dagger\hat{b}_j$ are, respectively, the boson creation, annihilation, and  number operators at the lattice site
$j$. The nearest-neighbor boson transfer amplitude is given by $t$; $U$ and $V$ parametrize the Coulomb repulsions between bosons resting at the same and neighboring sites. While $t$
causes the bosons to delocalize, promoting a superfluid (SF) phase  at weak interactions, $U$ $(V)$ tends to stabilize a MI (DW) when the interaction dominates over the kinetic energy scale set by $t$. 

In the framework of the DMRG the entanglement properties of the EBHM   
can be analyzed as follows. Consider the reduced density matrix 
$\rho_\ell=\mathrm{Tr}_{L-\ell}[\rho]$ of a block of length $\ell$ out
of a periodic system of size $L$. Then 
the bipartite entanglement spectrum $\{\xi_\alpha\}$ is defined 
as those of a fictitious  Hamiltonian ${\cal \bar{H}}$ defined via $\rho_{\ell}=e^{-{\cal \bar{H}}}$. 
As a consequence the $\xi_\alpha$ can be
extracted from the weights $\lambda_\alpha$ of 
the reduced density matrix $\rho_\ell$ by 
$\xi_\alpha=-2\ln\lambda_\alpha$.
Adding up, along the calculations, the $\lambda_\alpha$, we have direct access 
to the von Neumann entropy,  $S_L(\ell)=-\mathrm{Tr}_\ell[\rho_\ell\ln\rho_\ell]$.
On the other hand, from conformal field theory~\cite{CC04} one has 
$ S_L(\ell)=\frac{c}{3}\ln
           \left[\frac{L}{\pi}\sin\left(\frac{\pi\ell}{L}\right)\right]+s_1$ 
with the non-universal constant $s_1$. Thus we can easily
determine the central charge $c$ by DMRG.  
Since the most precise data for $S_L(\ell)$ were obtained when the length $\ell$ of the 
subblock equals half the system size $L$, the central charge should be determined 
from the relation~\cite{Ni11} 
\begin{eqnarray}
 c^\ast(L)= \frac{3[S_L(L/2-1)-S_L(L/2)]}{\ln[\cos(\pi/L)]}\;,
\end{eqnarray}
rather than directly using the above expression for  $S_L(\ell)$.

In contrast to hitherto existing  open boundary DMRG studies of the EBHM~\cite{DBA06,BDGA08,RF12}
we use periodic boundary conditions (PBCs). 
As shown for the regular Bose-Hubbard model this is advantageous calculating the central charge~\cite{EFGMKEvdL12,EBHEFS11}. Beyond that we benefit from the fact that no artificial on-site potentials at the edges
will affect  our results. To reach the same system sizes  as with open
boundary conditions (OBCs), we limit  the number 
of bosons per site. Throughout this work we use $n_b=2$; here the EBHM corresponds to an effective spin-1
Heisenberg model. We have convinced ourselves that at sufficiently large $U$ the boson truncation does not alter 
qualitatively the results presented in the following (solely, in the weak coupling regime, the extension of the SF phase
is somewhat underestimated). Let us finally note that we keep up to $m=2400$  states 
in the DMRG runs, so that the discarded weight is typically smaller than $1\times10^{-8}$.
For the dynamical DMRG calculations we take $m=800$ states to compute  the ground state during the first five DMRG sweeps, and 
afterwards use 400 states evaluating  the dynamical properties.

\begin{figure}[t]
 \begin{center}
  \includegraphics[clip,width=0.9\columnwidth]{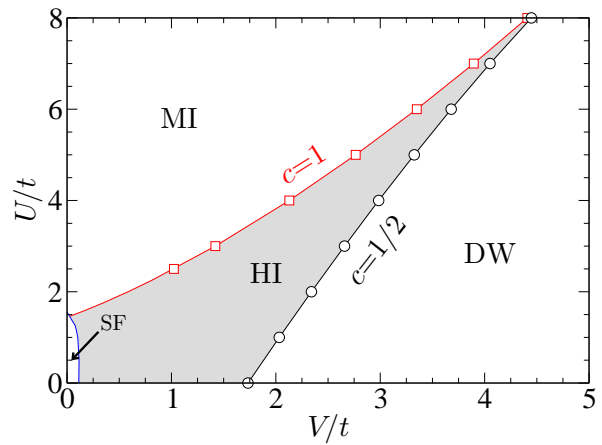}
 \end{center}
 \caption{(color online). DMRG  phase diagram of the 1D constrained extended Bose-Hubbard 
 model with  $n_b=2$ and $\rho=1$.  Shown are  the Mott insulator (MI), Haldane insulator (HI), density wave (DW), and superfluid (SF) phases. 
 The MI-HI (squares) and HI-DW (circles) transition points are determined via the central charge
 $c=1$ and $c=1/2$, respectively, which can be extracted from the von
 Neumann entropy [cf. Fig.~\ref{vN-gaps}~(b)]. 
 MI-HI transition points are confirmed by a  finite-size
 scaling of the two lowest energy levels with APBCs. 
Relaxing  the boson constraint the SF region extends.}
\label{pd}
\end{figure}

As stated above the ground-state phase diagram of the EBHM~\eqref{hamil} with $n_b=2$ 
exhibits three differing insulator phases, as well as a superfluid state at weak interactions
$U/t$, $V/t$. The stability regions of the various phases are pinpointed by Fig.~\ref{pd}. 
Let us emphasize that in the intermediate-coupling region ($3\lesssim U\lesssim 8$), the central charge
is best suited for detecting the MI-HI (HI-DW) quantum phase transition since the system
becomes critical at the transition points with $c=1$~(1/2).
\begin{figure}[t]
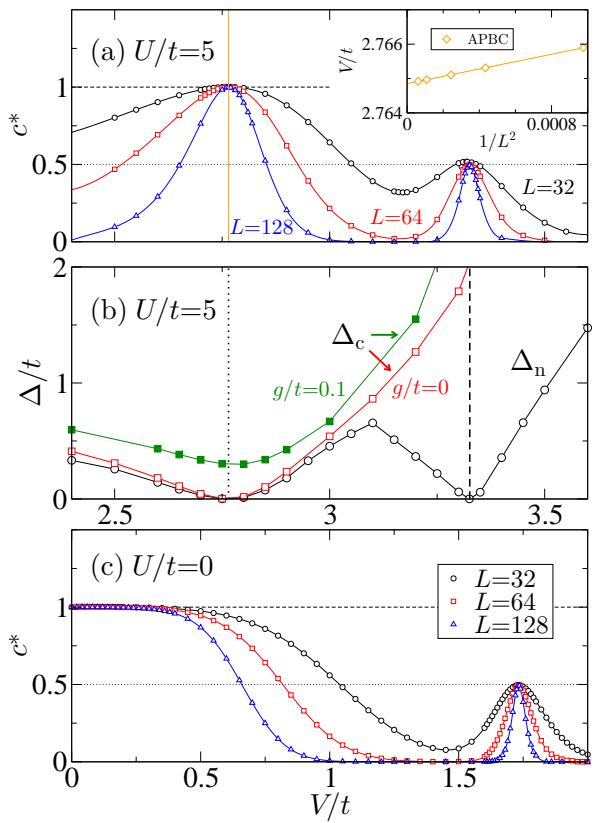

 \begin{center}
  \includegraphics[clip,width=0.9\columnwidth]{fig2a.eps}
  \includegraphics[clip,width=0.9\columnwidth]{fig2b.eps}
  \includegraphics[clip,width=0.9\columnwidth]{fig2c.eps}
 \end{center}
 \caption{(color online). Panel (a): Central charge $c^\ast$ of the EBHM with $U/t=5$, 
 indicating  the MI-HI (HI-DW) transition point with $c=1$ ($c=1/2$).
 The inset shows a finite-size scaling of the MI-HI transition points
 from the energy difference with APBCs.
 Panel (b): Extrapolated data for the charge gap $\Delta_c$ (open squares)
 and neutral gap $\Delta_n$ (open circles)  at $U=5t$.
 Vertical lines mark the transition points estimated from $c^\ast$.
 While $\Delta_n$ vanishes at both MI-HI and HI-DW boundaries,
 the charge gap $\Delta_c$ closes at the MI-HI transition only. 
 Turning-on an  inversion-symmetry breaking perturbation [$g/t=0.1$, see Eq.~(\ref{perturb-term})]
 $\Delta_c$ stays finite $\forall V/t$  (filled squares). 
 Panel (c): $c^\ast$ at $U=0$. Now the SF/MI-HI transition point is hardly to detect.}
\label{vN-gaps}
\end{figure}

Figure \ref{vN-gaps}~(a) illustrates the behavior of the central charge $c^*$ obtained numerically as a function of $V/t$ at fixed $U/t=5$.
With increasing system size $L$ two sharp peaks develop, indicating the MI-HI and HI-DW transition points. 
For $L=128$, we found $c^\ast\simeq 0.999$ in the former case and $c^\ast\simeq 0.494$ in the latter case, i.e., the numerical error, 
$|c^\ast(L)-c|/c$, is about 1\% if compared with the field theoretical predictions. Since the positions of the peaks only weakly depend  
on the system size, the transition points can be determined by
extrapolating the values of the critical $V(L)$ to the thermodynamic
limit $L\to \infty$. 
MI-HI transition points are also extracted from the level spectroscopy
of two lowest-lying energies with anti-periodic boundary conditions (APBCs),
$\hat{b}_{L+1}^{(\dagger)}\to-\hat{b}_{1}^{(\dagger)}$. This equates 
to the twisted boundary methods~\cite{KNO96} with the spin operators 
$\hat{S}_{L+1}^x\to-\hat{S}_{1}^x$ and $\hat{S}_{L+1}^y\to-\hat{S}_{1}^y$
applied to the spin-1 XXZ chain~\cite{CHS03}, see also Ref.~\cite{SM}. 
The obtained transition points can be linearly extrapolated to the
thermodynamic limit as in the inset of Fig.~\ref{vN-gaps}(a), 
showing a perfect agreement with the critical points obtained 
in the main panel.

The excitation gaps behave differently in various 
insulating phases~\cite{DBA06,BDGA08}: While the single-particle gap 
%\begin{equation}
$\Delta_c=E_0(N+1)+E_0(N-1)-2E_0(N)$
%\end{equation}
is finite in all three insulator phases, except for the MI-HI
transition point, the neutral gap 
%\begin{equation}
$\Delta_n=E_1(N)-E_0(N)$ 
%\end{equation}
closes both at the MI-HI and HI-DW transitions [$E_0(N)$ and $E_1(N)$ denote the energies of the
ground state  and first excited state of the $N$-particle system, respectively].
This is corroborated by Fig.~\ref{vN-gaps}~(b). 
A similar behavior of the neutral gap has been observed 
for the SPT phases of spin-1/2 ladder 
systems~\cite{MSHRG13}.
Note that the phase boundaries obtained by our PBC DMRG calculation 
at intermediate and strong couplings basically agree with very recent DMRG data for OBCs~\cite{RF12,RGGF13}. 
In the weak-coupling regime, on the other hand, our phase diagram differs  from former studies due to the 
 $n_b=2$ restraint. Accordingly the MI-SF transition at $V=0$ occurs  at a smaller value, $U\simeq 1.555t$, if 
compared to the critical $U/t$ derived from the Tomonaga-Luttinger liquid parameter~\cite{EFG11}.
The appearance of the SF phase, which can be understood as a Luttinger liquid with $c=1$~\cite{Gi03},
together with strong finite-size effects prevents using  $c^\ast$ for detecting the MI-HI  transition in this regime.
Otherwise, as shown by Fig.~\ref{vN-gaps}~(c), the HI-DW Ising transition  can still be determined from $c^*$, even for $U=0$. 

On these grounds, discussing the entanglement properties of the
SPT state, we consider the 
intermediate-coupling region hereafter. Calculating the entanglement spectrum $\xi_\alpha$ we divide the system in halves. Then, using DMRG with PBCs, 
one of the block with $L/2$ sites possesses two edges (rather than a single edge in the semi-infinite chain used by the infinite-time evolving 
block-decimation algorithm~\cite{PBTO12}).  
\begin{figure}[t]
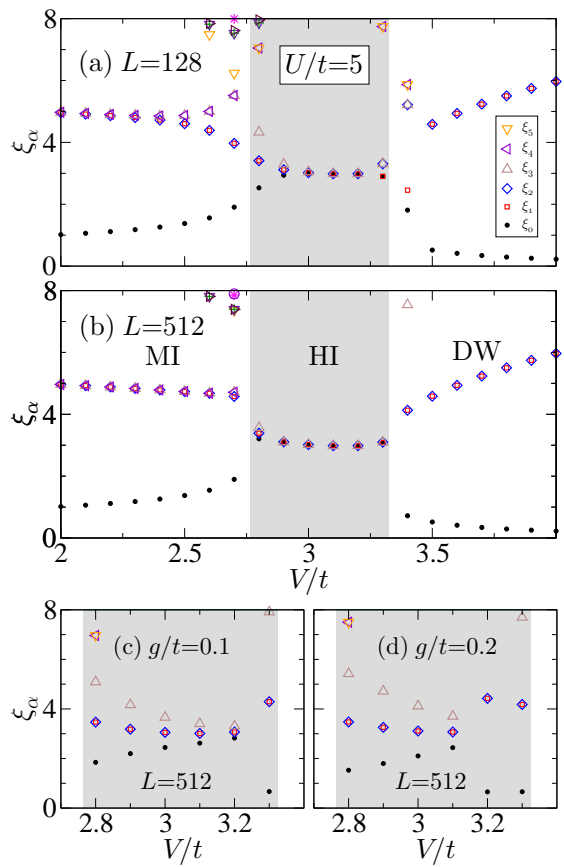

 \begin{center}
  \includegraphics[clip,width=0.85\columnwidth]{fig3a.eps}
  \includegraphics[clip,width=0.85\columnwidth]{fig3b.eps}
  \includegraphics[clip,width=0.85\columnwidth]{fig3c.eps}
 \end{center}
 \caption{(color online).
 Entanglement spectrum $\xi_\alpha$ of the EBHM with $U/t=5$.  
 %Again the dotted (dashed) lines mark the MI-HI (HI-DW) transition extracted from Fig.~\ref{vN-gaps}~(a). 
 If exciting the degeneracy of the entanglement levels becomes more perfect as the 
 system size increases (cf. data for $L=128$ [panel (a)] with those for $512$ [panel (b)]). 
 A perturbation~\eqref{perturb-term}  breaking the lattice inversion symmetry lifts the degeneracy in the HI phase.
 This is demonstrated by panels (c) and (d) giving  $\xi_\alpha$ for PBCs in the primary HI regime for $g/t=0.1$
 and $0.2$, respectively. 
 }
\label{ES-U5}
\end{figure}
In the HI phase the entanglement spectrum is expected to be at least four-fold degenerate, 
reflecting the broken $\mathbb{Z}_2\times\mathbb{Z}_2$ symmetry. 
Figure~\ref{ES-U5} shows the DMRG data for $\xi_\alpha$ obtained at $U/t=5$.
While for $L=128$ the four-fold degeneracy can be seen only deep inside of the HI phase,
for $L=512$ almost all HI states exhibit this degeneracy. By contrast, in the trivial MI and DW phases the lowest
entanglement level is always non-degenerate. Obviously higher entanglement levels $\xi_\alpha>8$ are also 
four-fold degenerate (cf. Fig.~\ref{EShigher-U5}  of Ref.~\cite{SM}). 
\begin{figure*}[t!]
 \begin{tabular}{ccc}
  \begin{minipage}{0.335\textwidth}
    \includegraphics[clip,height=5.7cm]{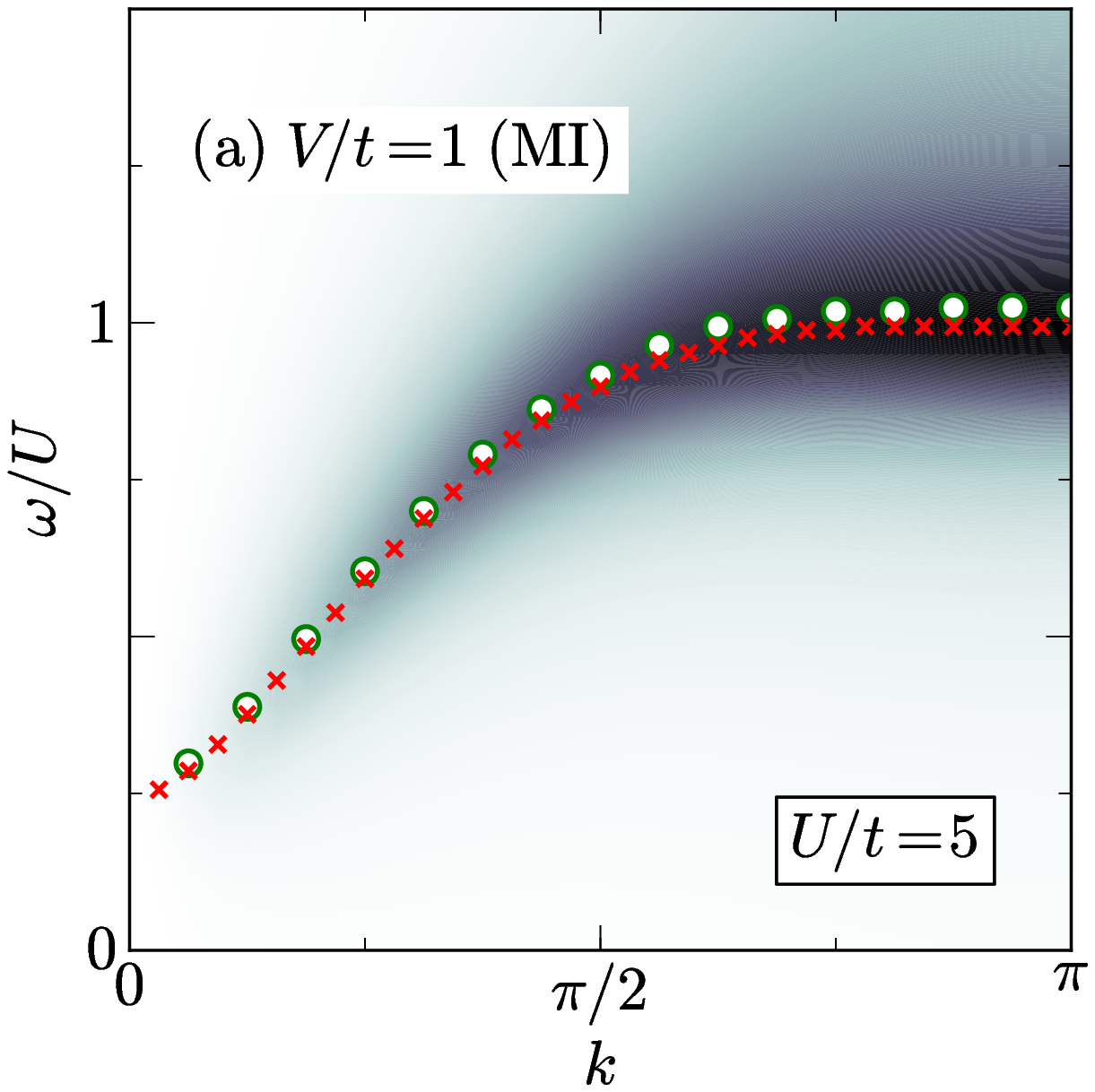}
  \end{minipage}
  \begin{minipage}{0.32\textwidth}
    \includegraphics[clip,height=5.7cm]{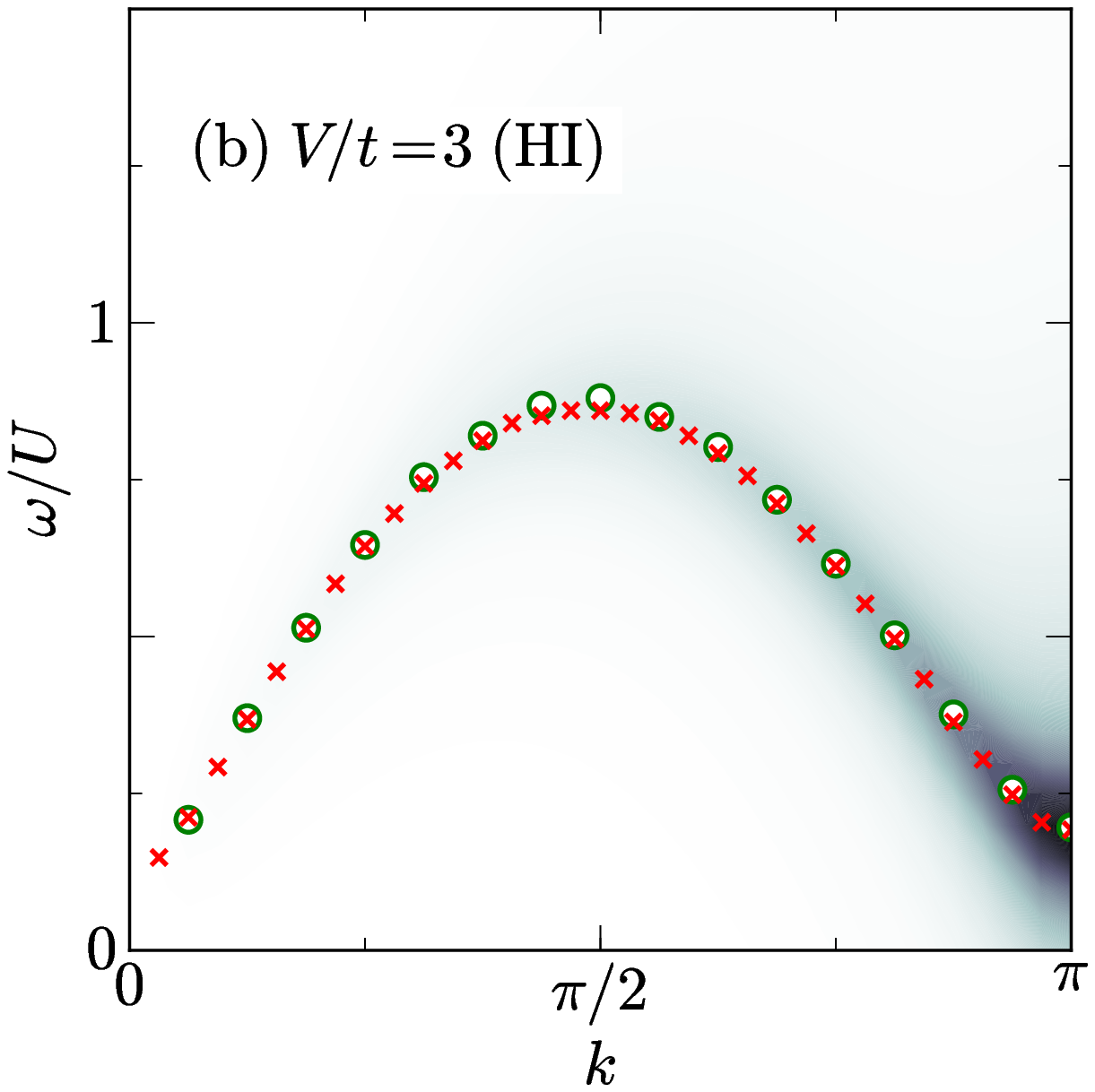}
  \end{minipage}
  \begin{minipage}{0.32\textwidth}
    \includegraphics[clip,height=5.7cm]{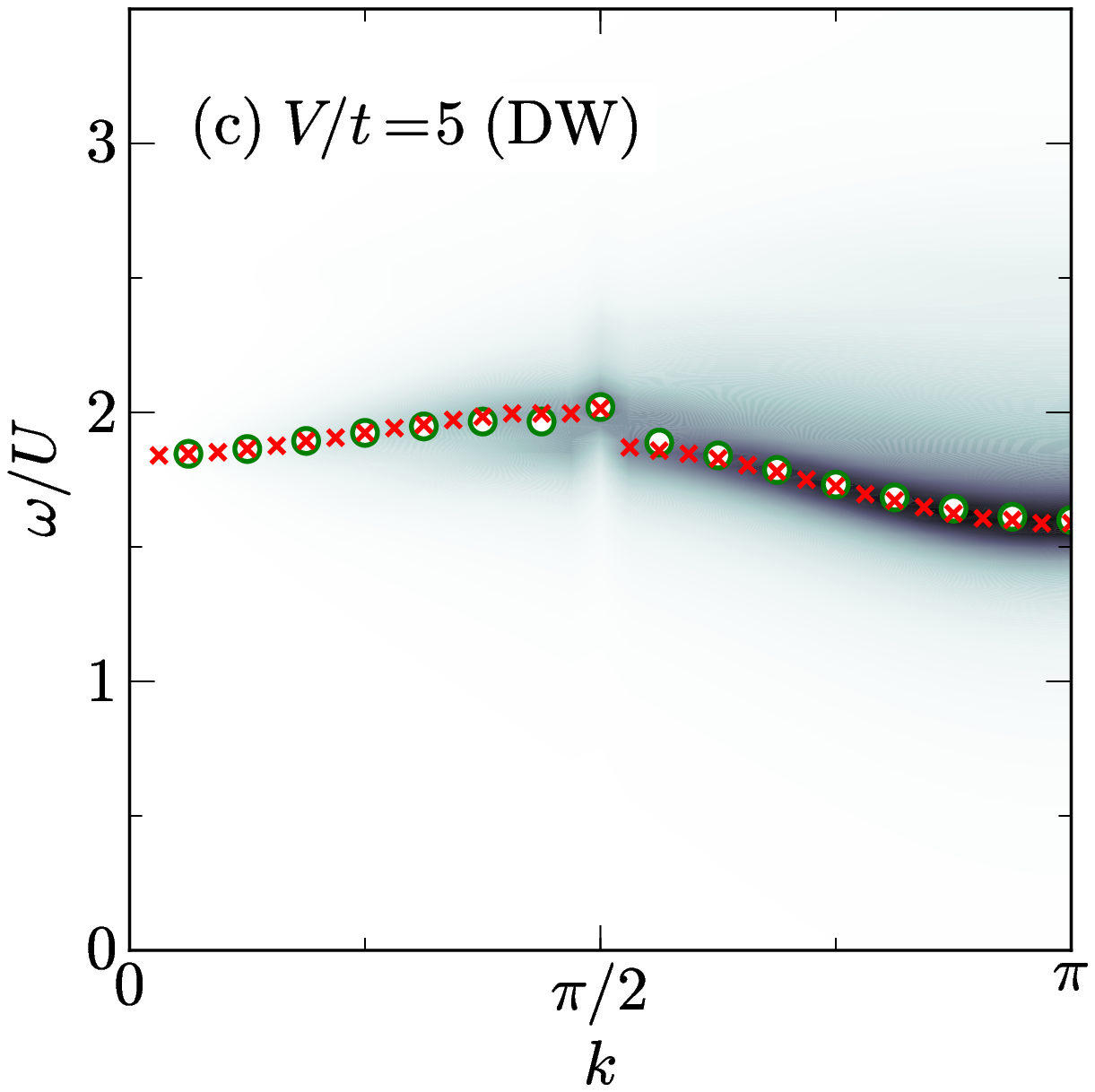}
  \end{minipage}
 \end{tabular}
 \caption{(color online).
 Intensity plots of the dynamical structure factor
 $S(k,\omega)$ in the MI (a), HI (b), and DW (c) phases. Data were obtained by  
 the dynamical DMRG technique for $L=64$ using a broadening $\eta=0.5t$.
 Crosses (circles) give the maximum value of $S(k,\omega)$ for
 $L=64$ ($L=32$ and $\eta=t$) at fixed momenta $k=2\pi j/L$ 
 with  $j=1,\,\cdots,\, L/2$.
 }
 \label{Skw}
\end{figure*}

%At this point we like to stress that the shaded region in the weak-coupling region of the phase diagram Fig.~\ref{pd}  can be 
%determined by detecting  just this four-fold degeneracy in the entanglement spectrum (cf., e.g., $\xi_\alpha$
%at $U=0$ shown in Fig.~\ref{ES-U0} of Ref~\cite{SM}).
 
We already stated that the HI phase is protected by the inversion symmetry of the lattice.
This symmetry can explicitly be broken by adding to the Hamiltonian~\eqref{hamil} an appropriate perturbation~\cite{BDGA08}:
\begin{eqnarray}
 \delta\hat{\cal H}
  = g \sum_j[(\hat{n}_j-\rho)
  \hat{b}_j^\dagger\hat{b}_{j+1}^{\phantom{}}+{\rm h.c.}]\,.
  \label{perturb-term}
\end{eqnarray}
As a consequence the MI-HI quantum phase transition   
disappears~\cite{BDGA08} and the single-particle charge gap  stays finite; see 
the filled squares in Fig.~\ref{vN-gaps}~(b) displaying $\Delta_c$ for $g/t=0.1$. 
One also expects that this perturbation lifts the degeneracy of the lowest 
entanglement level in the HI phase. Indeed Fig.~\ref{ES-U5}~(c) illustrates  
that any finite $g$ dissolves the four-fold degeneracy in the HI phase,
where the gap between the lowest levels  increases raising $g$ [cf. Fig.~\ref{ES-U5}~(d)]. 
That is, the entanglement spectrum substantiates the suspicion that
the lattice inversion symmetry is necessary for the nontrivial topological HI state to exist.

Since the EBHM~\eqref{hamil} can be realized by ultracold bosonic atoms loaded in optical lattices~\cite{BDZ08} 
it is highly desirable to study dynamical correlation functions which are accessible by experiments. For this purpose,  
the kinetic-energy correlations of the effective spin-1 Heisenberg chain was proposed to be a candidate detecting 
the HI phase and calculated on a mean-field level of approximation~\cite{DBA06}. 
Here we suggest the dynamical structure factor---which can be directly measured by momentum-resolved Bragg 
spectroscopy~\cite{CFFFI09,EGKPLPS09}---to be indicative of a  SPT state.
This quantity is defined as
\begin{eqnarray}
 S(k,\omega)=\sum_n |\langle \psi_n|\hat{n}_k|\psi_0\rangle|^2
             \delta(\omega-\omega_n)\,, 
\end{eqnarray}
where $|\psi_0\rangle$ and $|\psi_n\rangle$ denote the ground state and 
$n$th excited state, respectively.  The corresponding excitation energy
is $\omega_n=E_n-E_0$. 
In the absence of the nearest-neighbor repulsion~$V$, 
$S(k,\omega)$ was intensively studied  by means of perturbative and dynamical DMRG 
techniques~\cite{EFGMKEvdL12,EFG12}. Taking $V$ into account, 
in the MI, a gap opens at $k=0$ and the spectral weight becomes concentrated in the region $k>\pi/2$, around $\omega/U\simeq 1$, just as for the standard Bose-Hubbard model.
This is exemplified for $U=5t$ and $V=t$ by  Fig.~\ref{Skw}~(a). The maximum in $S(k,\omega)$ follows
a cosine-dispersion which is flattened, however, near the Brillouin zone boundary for $k\geq 3\pi/4$.
The situation dramatically changes when we enter the HI phase by increasing $V/t$, cf. 
Fig.~\ref{Skw}~(b) for $V/t=3$. Now the dispersion of the maximum in $S(k,\omega)$ bends back above $k=\pi/2$, acquiring a sinus shape
with (small) excitation gaps at both $k=0$ and $k=\pi$. Also the spectral weight of the dynamical charge structure factor is concentrated at 
$k=\pi$ and finite but very small for $\omega\ll U$. We note that the dispersion of the maximum in the HI phase is remindful of those of the spin-1 Heisenberg chain. 
A dispersive signal persists if we allow larger $n_b$ (see the results presented in
Ref.~\cite{SM} for the EBHM with   $n_b=5$). In the DW phase, the maximum 
of $S(k,\omega)$ is almost dispersionsless and located at $\omega \gtrsim 1.5 U$ for $U/t=V/t=5$ [see Fig.~\ref{Skw}~(c)].  The intensity is notably 
more confined than for the MI.  Figure~\ref{Skw} demonstrates that the dispersion in the insulating phases barely changes if the system
size is increased. 
In every sense, $S(k,\omega)$ behaves very differently in the  MI, DW, and HI states and might therefore
be used to discriminate these insulating phases.  
   
In summary,  we studied---from an entanglement point of view---the topologically nontrivial Haldane insulator, appearing in the intermediate coupling regime of the 
1D Bose-Hubbard model with on-site and nearest-neighbor Coulomb interactions in the midst of Mott insulator, 
density-wave  and  superfluid phases. Using the  DMRG 
technique,  the MI-HI (HI-DW) quantum phase transition is  determined with high precision from the central charge $c^\ast$ 
that can be extracted from the von Neumann entropy. We thereby approved the universality class $c=1$ ($c=1/2$) predicted by field theory.
We  furthermore established a characteristic four-fold degeneracy of the lowest entanglement level in the SPT Haldane 
phase and demonstrated that any violation of the lattice inversion symmetry lifts this degeneracy. With the objective to stimulate further experiments on ultracold
bosonic atoms in optical lattices  we analyzed the dynamical charge structure factor  for the extended Bose-Hubbard model and 
showed that this quantity can be used to distinguish the Haldane insulator,  exhibiting a gapped excitation spectrum similar to the spin-1 Heisenberg-chain model, 
from conventional Mott and density-wave states. 

% \section*{Acknowledgment}
{\it Acknowledgments.} The authors would like to thank S.~Nishimoto and T.~Yoshida
for valuable discussions. This work was supported by Deutsche Forschungsgemeinschaft 
through SFB 652, Project B5.

%%%%%%%% References  %%%%%%%%%%%%%%%%%%%%%%%%%%%%%%%%%%%%%%%%%%%%%%
%\bibliography{Holger,ref-ebhm,ref-ebhm2}
\bibliographystyle{apsrev4-1}
%

%%%%%%%%%SUPPLEMENTARY MATERIAL%%%%%%%%%
\clearpage
\appendix 
\section{\Large 
  Supplementary material}
\renewcommand{\theequation}{$\text{S}$\arabic{equation}}
\setcounter{equation}{0}
\renewcommand{\thefigure}{$\text{S}$\arabic{figure}}
\setcounter{figure}{0}
Using the unbiased density matrix renormalization group (DMRG) technique
with periodic boundary conditions (PBCs),
in the main paper,  we derived the ground-state phase diagram of
the one-dimensional (1D) extended Bose-Hubbard model (EHBM), 
restricting the maximum number of bosons per site to be $n_b=2$. 
Figure~\ref{pd} showed the extent of the Haldane insulator (HI) phase, located  
between the conventional insulating Mott  (MI) and density wave (DW)
states and the superfluid (SF) phase.  
In the following we provide further results for the EBHM in the intermediate-coupling region, in order to guarantee  that 
our main conclusions  will be unaffected in the more general case with $n_b>2$.

\subsection{Determination of the MI-HI transition points}

\begin{figure}[b]
 \begin{center}
  \includegraphics[clip,width=\columnwidth]{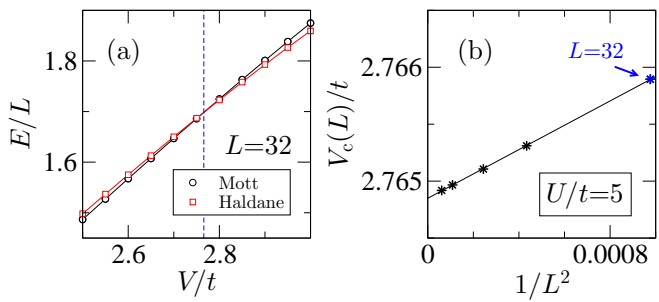}
 \end{center}
 \caption{
 (a) The $V$ dependence of the two lowest energy eigenvalues with
APBCs  at $U/t=5$ and $L=32$. The energies
 of the Haldane state (squares) and the Mott insulating state (circles)
 cross at the MI-HI transition point. (b) The critical points 
 $V_{\rm c}(L)/t$ as extracted in the panel (a) versus inverse
 of the squares of the system size at $U/t=5$ with up to $L=128$. 
 }
\label{twistBC_Vc}
\end{figure}

As stressed in the main text the 1D EBHM can be mapped onto an  effective spin-1 XXZ
chain model with on-site anisotropy $D$. Then, in the notations of Ref.~\cite{CHS03}, 
the three insulating MI, HI, and DW phases of the EBHM 
correspond to the large-$D$, the Haldane and the N{\'e}el phases,
respectively. The Haldane phase in the spin-1 chain is described by a spin-1/2 two-leg ladder 
system~\cite{PTBO10,PBTO12,LYHYW12}. As discussed in Ref.~\cite{LYHYW12},  with the projective
representations of the symmetry group it is related to the $t_0$ phase of the latter system.
According to Ref.~\cite{CHS03}, using Lanczos diagonalization, the large-$D$-Haldane phase transition points can be determined
by level spectroscopy~\cite{KNO96} of the two lowest-lying energies with 
twisted boundary conditions (i.e., $\hat{S}_{L+1}^x\to-\hat{S}_{1}^x$, 
$\hat{S}_{L+1}^y\to-\hat{S}_{1}^y$, and $\hat{S}_{L+1}^z\to\hat{S}_{1}^z$). 
Hence also the MI-HI transition points can be extracted by analyzing  
the two lowest-lying energies with anti-periodic boundary conditions (APBCs),
i.e., $\hat{b}_{L+1}^{(\dagger)}\to-\hat{b}_{1}^{(\dagger)}$. 
As shown by Fig.~\ref{twistBC_Vc}~(a), the Mott insulating state and 
the Haldane state crosses at the MI-HI transition point $V_{\rm c}(L)$
for the fixed system size used at $U/t=5$. The transition points obtained 
for various system sizes can be linearly extrapolated to the
thermodynamic limit $L\to\infty$, see Fig.~\ref{twistBC_Vc}(b).  
We emphasize the perfect agreement with the critical points obtained 
in the main panel of Fig.~\ref{vN-gaps}(a).

\subsection{Entanglement spectrum in the constrained EBHM}
\begin{figure}[tb]
 \begin{center}
  \includegraphics[clip,width=0.9\columnwidth]{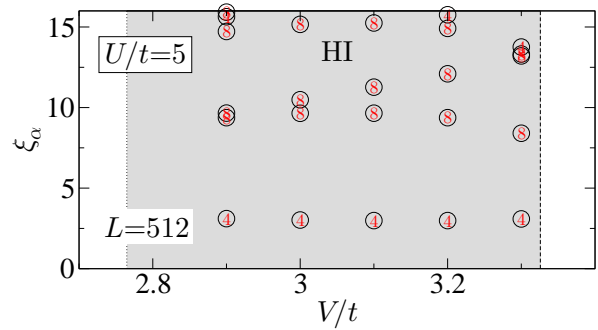}
 \end{center}
 \caption{
 Entanglement spectrum $\xi_\alpha$ obtained by DMRG in the HI state of the EBHM with $U/t=5$. 
 The dotted (dashed) line denotes the MI-HI (HI-DW) transition point extracted from the von Neumann entropy, 
 see Fig.~\ref{vN-gaps}(a) of the main text. The numbers in the circles
 give the degree of degeneracy.
 }
\label{EShigher-U5}
\end{figure}
In the HI, due to the broken 
$\mathbb{Z}_2\times\mathbb{Z}_2$ symmetry, 
the lowest entanglement level is four-fold degenerate Since the HI phase is
nontrivial protected by the lattice inversion symmetry, not only 
the lowest but also the entire entanglement spectrum is $4j$-fold 
degenerate with $j=1,2,\ldots$  Figure~\ref{EShigher-U5} 
visualizes the four-fold degeneracy of the higher entanglement levels
in the HI phase at $U/t=5$ for a system with $L=512$, PBC and $n_b=2$. 
 
\begin{figure}[tb]
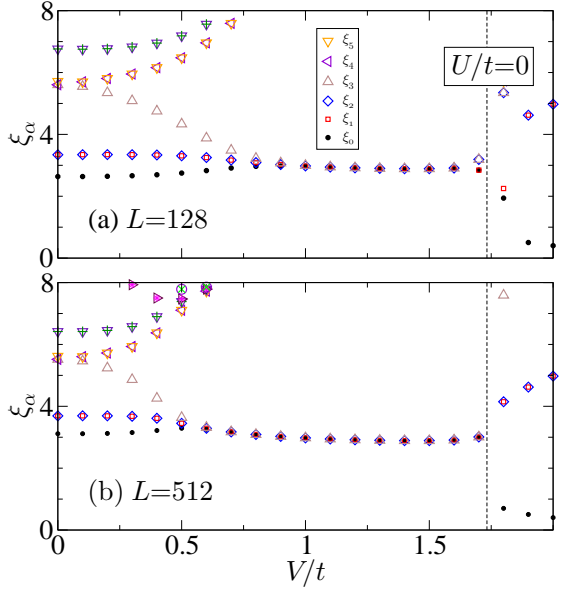

 \begin{center}
  \includegraphics[clip,width=0.85\columnwidth]{figS3a.eps}
  \includegraphics[clip,width=0.85\columnwidth]{figS3b.eps}
 \end{center}
 \caption{
 Entanglement spectrum $\xi_\alpha$ in the weak-coupling  ($V/t$)  
 regime at $U/t=0$. Data obtained by DMRG with PBCs for $L=128$ (a) and $L=512$ (b). 
 The dashed line gives the HI-DW transition point extracted
 from the von Neumann entropy.
 }
\label{ES-U0}
\end{figure}

As noticed in the main text, in the weak-coupling regime the  central charge $c^\ast(L)$  strongly depends 
on the system size  [cf. Fig.~\ref{vN-gaps}~(c)]. 
Then the MI-HI (and likewise the SF-HI) phase transition  is hard to detect. 
Figure~\ref{ES-U0} demonstrates that the situation is similar for the entanglement spectrum 
calculations at $U=0$ with $n_b=2$. Close to the Ising transition point ($U\approx1.733$) 
the lowest entanglement level is four-fold degenerate, 
indicating the existence of the HI phase also at $U=0$. 
Increasing the system size the HI state extends to $V\to0$.  
Here a more precise finite-size-scaling is desired to pinpoint the MI/SF-HI
transition. 
%Note that the SF phase is underestimated because of the limitation
%$n_b=2$, while the tendency of the system size dependence in the 
%weak-coupling regime for $n_b>2$ should be similar.

%%%%% n_b=5  %%%%
\subsection{Haldane insulator state in the full EBHM}
\begin{figure}[b]
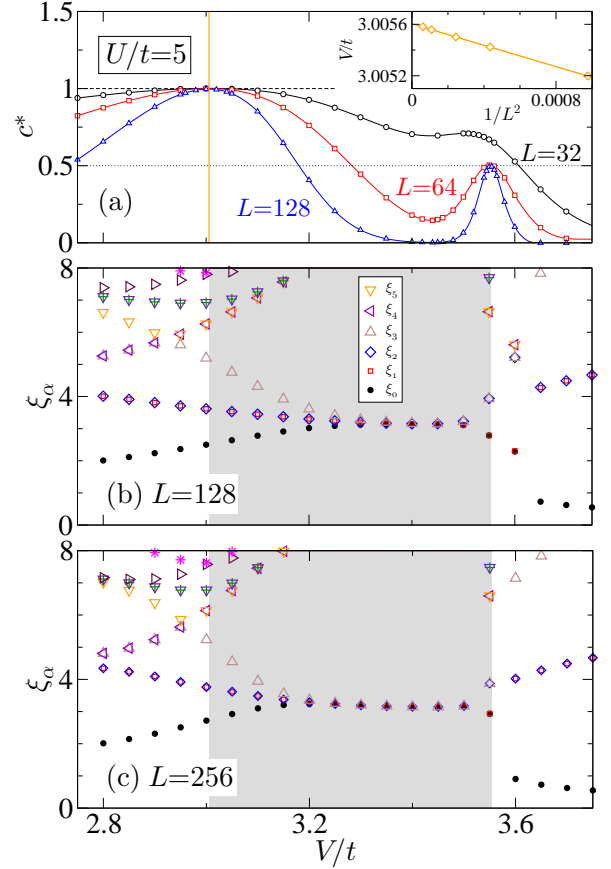

 \begin{center}
  \includegraphics[clip,width=0.9\columnwidth]{figS4a.eps}
  \includegraphics[clip,width=0.9\columnwidth]{figS4b.eps}
  \includegraphics[clip,width=0.9\columnwidth]{figS4c.eps}
 \end{center}
 \caption{
 (a) Central charge $c^\ast(L)$ at $U/t=5$ with $n_b=5$,
 indicating the MI-HI (HI-DW) transition points with $c=1$ ($c=1/2$). 
  $\xi_\alpha$ in the 1D EBHM with $U/t=5$ for $n_b=5$ and $L=128$ (b) respectively  $L=256$ (c).
 }
\label{cc-U5-nb5}
\end{figure}

We now demonstrate that the qualitative analysis of EBHM with 
$n_b=2$  remains valid if we increase the boson cutoff.
To this end we convinced ourselves that for large enough values of $U/t$ 
again both MI-HI and HI-DW phase
transition points can be determined via the entanglement entropy
and the level spectroscopy.
For example, at $U=5t$, the MI-HI phase transition occurs
at $V/t\simeq 3.00$ with $c^\ast\simeq 1.0$ and the system-size 
dependence of maxima is very weak; see main panel of
Fig.~\ref{cc-U5-nb5}(a). 
Adopting level spectroscopy  again, the MI-HI transition points can be determined 
(see inset), yielding excellent agreement with the values in the main panel.
The Ising-like HI-DW transition shows up for large system sizes $L>32$ 
at $V/t\simeq 3.55$ with $c^\ast\simeq 0.5$. 
The entanglement spectrum $\xi_\alpha$ with $n_b=5$ shows 
a degeneracy of the lowest level deep in the HI phase for $L=128$ 
as in Fig.~\ref{cc-U5-nb5}(b).
With increasing system size the degenerate HI state extends to
the point of the MI-HI transition [Fig.~\ref{cc-U5-nb5}(c)].

\begin{figure}[htb]
 \begin{center}
  \includegraphics[clip,height=5.7cm]{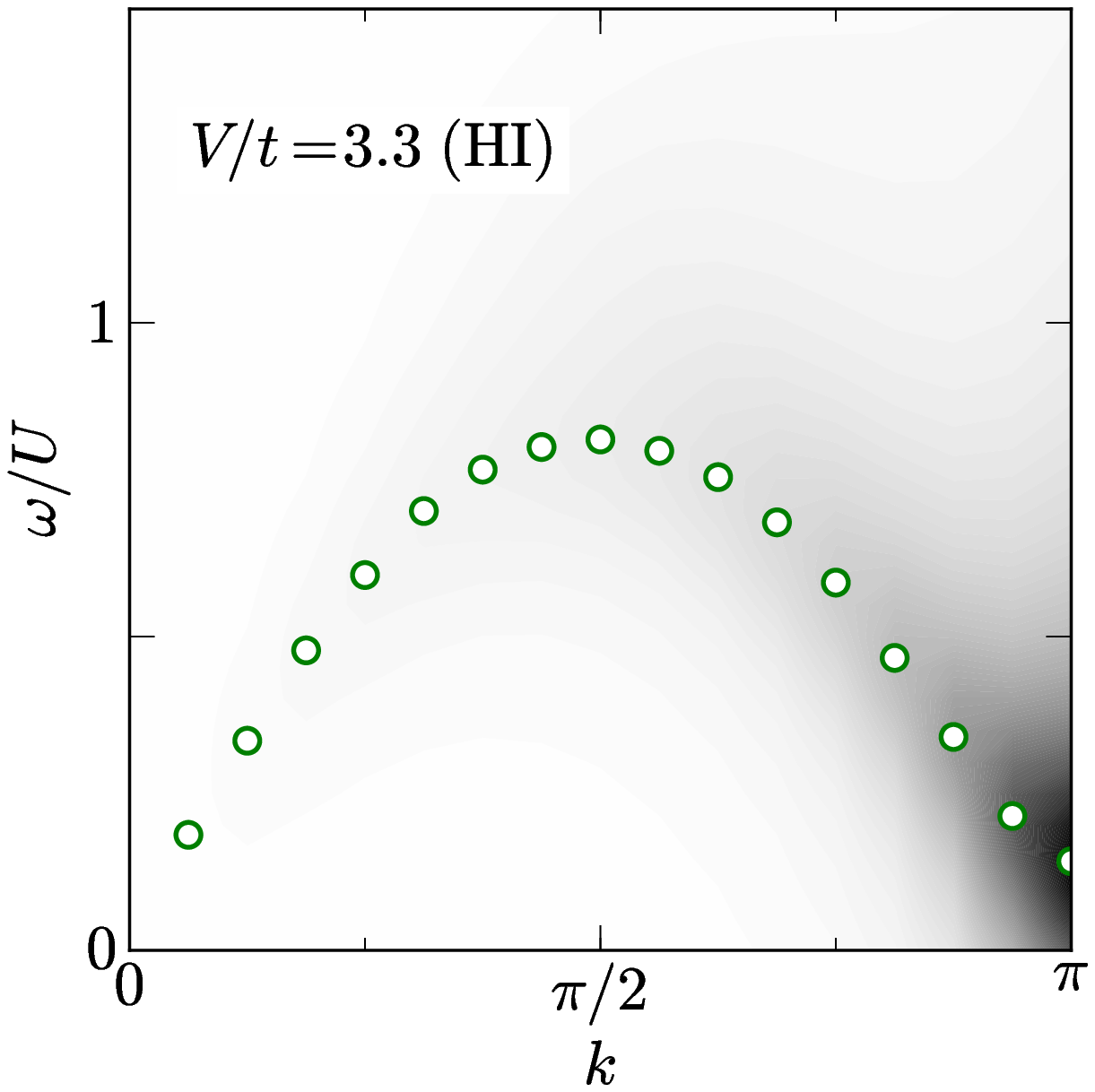}
 \end{center}
 \caption{
 Intensity plot of $S(k,\omega)$ in the EBHM with cutoff $n_b=5$.
 Results were obtained for a lattice with $L=32$ sites and PBCs, 
 where $U/t=5$ and $V/t=3.3$.
 Within the dynamical DMRG a broadening $\eta/t=1$ is used. Circles give the
 maxima in $S(k,\omega)$ for $k=2\pi j/L$ where $j=1,\,\cdots,\, L/2$.
 }
\label{HI-Skw-nb5}
\end{figure}

Finally we show that characteristic behavior of the dynamical 
charge structure factor $S(k,\omega)$ in the HI phase
survives the inclusion of higher boson occupation numbers. 
Figure~\ref{HI-Skw-nb5} presents dynamical DMRG results 
for $S(k,\omega)$ with $n_b=5$ deep in the HI phase (for $U/t=5$,  $V/t=3.3$,  
$L=32$ and broadening $\eta=t$). Just as for the constrained
EBHM with $n_b=2$, most of the spectral weights in $S(k,\omega)$ 
is concentrated  around $k=\pi$ and $\omega\ll U$. The maxima of $S(k,\omega)$ 
follow---as a function of the momentum---the sinus-like dispersion known from the dynamical
spin structure factor in the quantum spin-1 Heisenberg model. 
Note that the system-size dependence of the dispersion in $S(k,\omega)$ is  hardly
seen in Fig.~\ref{Skw}. 
We conclude that our results for the dynamical structure factor 
in the HI phase of the constrained EBHM hold qualitatively in the full EBHM as well.

%%%%%%%%%END OF SUPPLEMENTARY MATERIAL%%%%%%%%%

\end{document}